\documentclass[aip,reprint,showpacs]{revtex4-1}

\draft 
\usepackage[utf8]{inputenc}
\usepackage{amsmath,amsthm,amssymb}
\usepackage{float}
\usepackage{graphicx}
\usepackage{dcolumn}
\usepackage{bm}
\usepackage{color}
\usepackage[caption=false]{subfig}
\usepackage{hyperref}   
\usepackage[all]{hypcap}
\hypersetup{
    colorlinks=true,
    linkcolor=blue,
    filecolor=magenta,
    urlcolor=cyan,
    citecolor=red
     }
\pdfoutput=1
\DeclareUnicodeCharacter{2212}{-}

\begin{document}
\title{Fermi acceleration in relativistic collisionless plasma shocks correlates with anisotropic energy gains}
\author{Roopendra Singh Rajawat}
\affiliation{School of Applied and Engineering Physics, Clark Hall, Cornell University, NY 14850, USA}
 \author{ Vladimir Khudik}
\affiliation{Department of Physics and Institute for Fusion studies, The University of Texas at Austin, TX 78712, USA}
\author{Gennady Shvets}
\affiliation{School of Applied and Engineering Physics, Clark Hall, Cornell University, NY 14850, USA}

\date{\today}
\begin{abstract}
Collisionless shocks generated by two colliding relativistic electron-positron plasma shells are studied using particle-in-cell (PIC) simulations. Shocks are mediated by the Weibel instability (WI), and the kinetic energy of the fastest accelerated particles is found to be anisotropically modified by WI-induced electric fields. Specifically, we show that all particles interacting with the shock bifurcate into two groups based on their final relativistic Lorentz factor $\gamma$: slow ($\gamma < \gamma_{\rm bf}$) and fast ($\gamma > \gamma_{\rm bf}$), where $\gamma_{\rm bf}$ is the bifurcation Lorentz factor that was found to be approximately twice the initial (upstream) Lorentz factor $\gamma_0$. We have found that the kinetic energies of the slow particles are equally affected by the longitudinal and transverse components of the shock electric field, whereas the fast particles are primarily accelerated by the transverse field component.
\end{abstract}

\pacs{52.27.Ep, 52.65.−y, 52.35.Tc}
\maketitle
\section{Introduction} \label{introduction}
Relativistic collisionless shocks are widely viewed as efficient sources of particle acceleration in blazars, supernova remnants and gamma-ray burst (GRB) afterglows \cite{Piran_rmp_2005}, as well as other high-energy astrophysical objects~\cite{Medvedev_1999}. Collisionless shocks are ubiquitous in low-density astrophysical plasmas, where energy is dissipated through effective collisions provided by particles' interactions with turbulent electromagnetic fields. In the absence of binary collisions, such effective collisions enable particle acceleration to ultra-relativistic energies. Understanding the emergence of the underlying electromagnetic turbulence via collective plasma instabilities is crucial for understanding the physics of particle acceleration in astrophysical contexts, where plasma densities are low and binary collisions can be mostly neglected. Moreover, the complexity of instability-mediated electromagnetic fields raises questions about the exact mechanism behind the acceleration of fast particles, as well as the factors distinguishing such minority populations from the majority of particles that never reach high energies.

In the specific case of relativistic unmagnetized plasma flows interacting with each other or with the interstellar medium (ISM) plasmas, the classic Weibel Instability~\cite{Weibel} (WI) is widely viewed as responsible for the spontaneous generation of sub-equipartition electromagnetic fields~\cite{Chang_2008, Keshet_apj_2009, Neda_pop_2018}, mediation of collisionless shocks, and generation of superthermal particles \cite{Silva_2003,Milosavljevi__2006,Spitkovsky_aip_2005,Spitkovsky_2008,Sironi_2013,Haugbolle_apj_2011} in various astrophysical scenarios. The WI is a collective electromagnetic instability which develops in plasmas with anisotropic velocity distributions. Analytical and simulation studies show that the WI \cite{Medvedev_1999,Silva_2003} generates large magnetic fields that can reach the Alfv\`{e}nic limit during its nonlinear stage \cite{Yoon_pra_1987, Kato_pop_2005,Polomarov_prl_2008,Shvets_pop_2009,Bret_pop_2010}.

Particle-in-cell (PIC) simulations have long been the primary tools for studying the effects of the WI-generated electromagnetic turbulence on the generation of superthermal particles. For example, power-law ({\it i.e.} non-Maxwellian) particle distributions $f(\varepsilon) \sim \varepsilon^{-p}$ as a function of the particle energy $\varepsilon$ have been predicted based on PIC simulations results, with consistent power-law coefficients $p$ have extracted by several groups, e.g., $p \simeq 2.4$ in two-dimensional (2D)~\cite{Spitkovsky_aip_2005,Spitkovsky_2008} and $p \simeq 2.1-2.3$ in three-dimensional (3D)~\cite{Haugbolle_apj_2011} geometries. Particle acceleration has been found to be governed by stochastic diffusion, where particles move back and forth across the shock front and gain energy by scattering from self-consistent magnetic turbulence through the first-order Fermi acceleration mechanism~\cite{Fermi_pr_1949}. Consistently with phase space diffusion mechanism of particle acceleration, the maximum energy gain of particles in Fermi acceleration is observed to scale with acceleration time ($t_{acc}$) according to\cite{Sironi_2013} $\epsilon_{\rm max} \propto t^{1/2}_{acc}$. In the aforementioned references, Fermi acceleration is identified by the existence of non-thermal tail of the distribution function and kinetic energy carried by the non-thermal particles.

Nevertheless, the detailed micro-physics behind particle acceleration is still poorly understood. Because of the complex vectorial nature of the electromagnetic turbulence inside the collisionless shock itself and in the shocked plasma, systematic tracking of the particles passing through the shock is needed to answer many specific questions. These include: (i) what is the relative role of different components of the electric field in particle acceleration? (ii) what distinguishes the majority of thermalized particles in the shocked region from the minority particles gaining most of the energy? (iii) what are the telltale signs of Fermi acceleration that can be extracted from such tracking? Note that while magnetic fields are dominant inside the shock, and are primarily responsible for particles' thermalizations, they can neither accelerate nor decelerate charged particles -- this is done by the much weaker electric fields. While particle tracking has been used in the past to track the accelerated particles~\cite{Spitkovsky_2008,Martin_2009,Plotnikov_mnras_2018}, it has not been used to investigate the highly-anisotropic nature of particle acceleration, as expressed by relative contributions of the longitudinal (parallel to the front velocity) and transverse components of the electric field to particles' energy gains/losses.

In this work, we present the details of numerical tracking of representative particles extracted from first-principles 2D PIC simulations of the relativistic, unmagnetized electron-positron (pair) plasma shocks. The 2D geometry is sufficient for capturing the basic physics of particle acceleration. The rest of the manuscript is organized as follows. In Section~\ref{method}, we describe the geometry and physical parameters of the problem at hand, and the details of the PIC simulation setup. The structure of the shock and the dynamics of the bulk plasma (pre-shock and shocked) are discussed in Section~\ref{structure}. A detailed balance between work done on accelerated particles by different electric field components is discussed in Section~\ref{bifurcation} and the acceleration process is described using particle tracking in Section \ref{single_particle_dynamics}. We describe the work-energy bifurcation, which reveals two distinct groups of particles. The particles from the first group gain energy in equal measure from the longitudinal and transverse electric fields, while those from the second group derive most of their energy from the transverse electric field. These distinct groups of particles indicate the difference in wave-particle interaction between bulk and superthermal plasmas. In Section \ref{sec:reflected_particles}, we discuss the role of the shock-reflected particles. The conclusions are presented in Section~\ref{conclusions}.

\section{Physical setup and simulation details}\label{method}
\begin{figure}[ht!]
\includegraphics[width=1\linewidth]{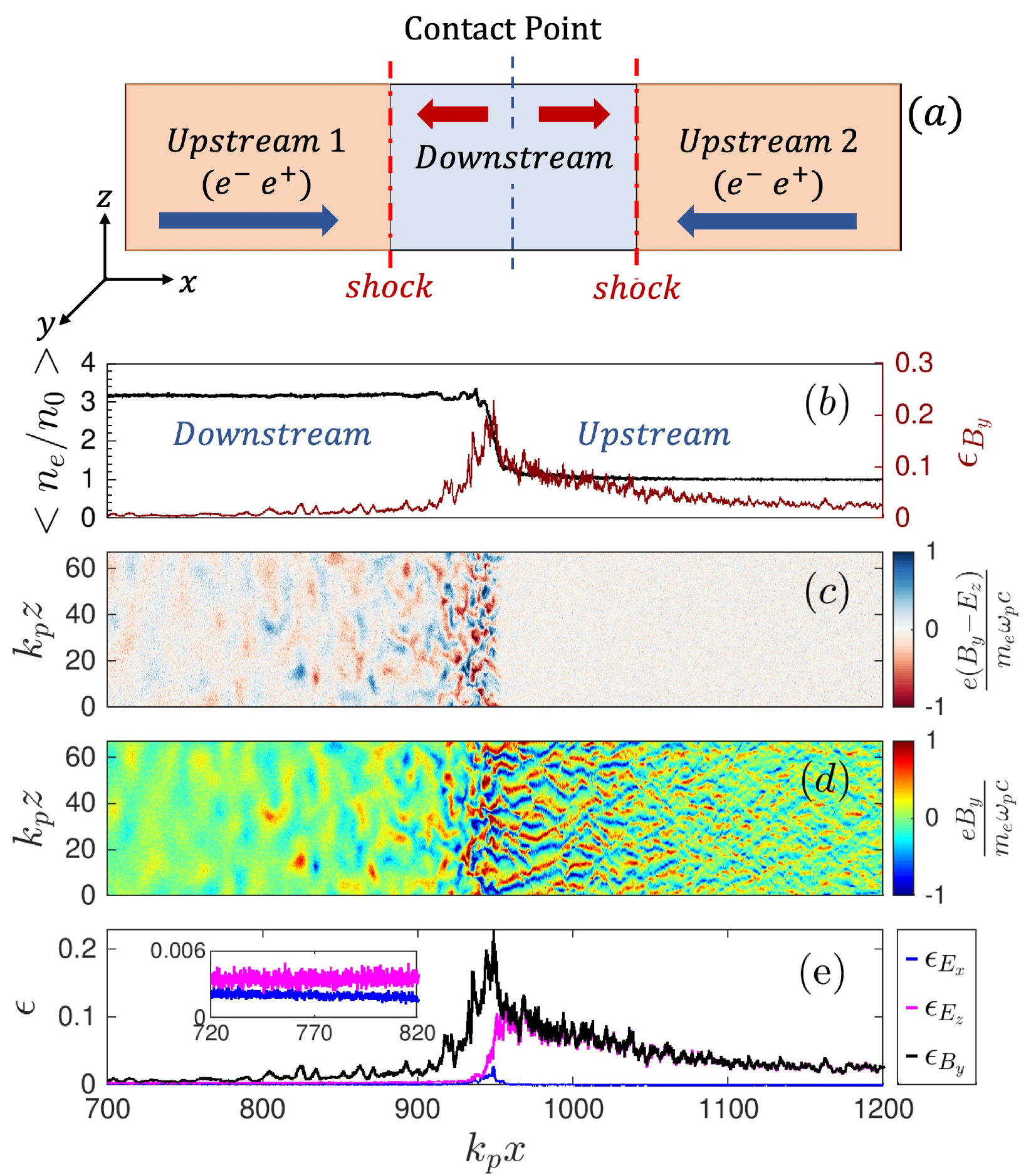}
\caption{(a) Two counter-propagating pair plasmas collide and produce two counter-propagating shocks. (b-e) Typical snapshots ($t = t_{\rm fin} \simeq 2040 \omega^{-1}_{p}$) of the shocked (``downstream'') and pre-shocked (``upstream'') plasmas. (b) Electron density, (c) magnetic field, (d) transversely averaged electron density, and (e) transversely averaged electromagnetic field energy densities: longitudinal electric $\epsilon_{x}^{(E)}$ (blue), transverse electric $ \epsilon_{z}^{(E)}$ (magenta), and transverse magnetic $\epsilon_{B}$ (black) lines. Inset: zoom into the $720 k_{p}^{-1} \leq x \leq  820 k_{p}^{-1}$ range.}
\label{fig:density_fieldenergy}
\end{figure}

The physical setup of the problem is schematically illustrated in Fig.~\ref{fig:density_fieldenergy}(a): two streams of cold electron-positron plasmas counter-propagate along the $x$-direction and come into the initial contact in the plane marked by a black dashed line. In the rest of the manuscript, we assume that the two electrically-neutral streams are mirror images of each other, and that their initial Lorentz factors and laboratory frame densities for each of the species are $\gamma_{0} = 20$ and $n_0$, respectively. The mirror symmetry enables a standard computationally-efficient approach~\cite{Kato_2007} to modeling colliding plasmas: the stream $2$ particles are reflected off a stationary wall placed in the plane of contact ($x=0$).  Perfectly-conducting boundary conditions for the electromagnetic fields are imposed at $x=0$.

Under this approach, unperturbed streaming plasma is continuously injected through the right boundary. As the plasma is reflected by the wall and collides with the incoming plasma, a region of counter-streaming plasma is formed in the overlapping region. The resulting extreme anisotropy of the mixed plasma triggers the WI and eventually leads to the formation of a collisionless shock (red dashed line in Fig.~\ref{fig:density_fieldenergy}(a)) propagating in the $+x$-direction. As the incoming cold plasma encounters strong electromagnetic fields in the shock region, it gets thermalized and forms an isotropic hot plasma in the shocked region. By symmetry, the plasma in this region behind the shock (further referred to as the downstream region) has a vanishing overall drift velocity. The simulation is carried out in the laboratory reference frame of a stationary reflective wall, where the downstream plasma is, on average, at rest.

The above described problem is numerically solved using a 2D ($y$-independent) version of a PIC code VLPL \cite{Pukhov_1999, Pukhov_jcp_2020}. A novel rhombi-in-plane scheme \cite{Pukhov_jcp_2020} is used for updating the electromagnetic fields, which is designed to suppress the numerical Cherenkov instability. The non-vanishing electromagnetic field components $B_y$ (out-of-plane), $E_x$ (longitudinal), and $E_z$ (transverse) are assumed to be functions of $(x,z)$ and $t$, and the only non-vanishing components of the electron/positron momenta are $p_x$ and $p_z$.

The natural scales for time and space are the inverse values of the plasma frequency $\omega_{p}^{-1}$ and wave number $k^{-1}_{p} (= c/\omega_{p})$, respectively. Here $\omega_{p} = \left( 4 \pi n_{0} e^{2}/\gamma_{0} m_{e}  \right)^{1/2}$ is the relativistic plasma frequency, $-e$ and $m_{e}$ are the electric charge and mass of an electron.
The size of the simulation domain is chosen to be $L_{x} \, \times \, L_{z} = 2100 k^{-1}_{p} \times 67 k^{-1}_{p} $. The following spatial grid cell $(\Delta x, \Delta z)$ and time step $\Delta t$ were used: $\Delta x = 0.07 k^{-1}_{p}$, $\Delta z  = 3 \Delta x$, and $\Delta t = \Delta x/c$~\cite{Pukhov_jcp_2020}. For all simulations, $16$ particles per grid cell per species were used.

\section{Review of the Shock Structure}\label{structure}
Below we review the well-established properties of the shocked (downstream) and pre-shocked (upstream) plasmas, as well as that of the shock created by the collision of counter-streaming plasmas \cite{Silva_2003,Milosavljevi__2006,Spitkovsky_aip_2005,Spitkovsky_2008,Keshet_apj_2009, Sironi_2013,Haugbolle_apj_2011,Lemoine_prl_2019,Lemoine_pre_2019a,Lemoine_pr_2019b,Lemoine_pre_2019c}. Unless stated otherwise, all figures are plotted at $\omega_{p}t_{\rm fin} \simeq 2040$ chosen to ensure that the shock region is well-formed.

A sharp density transition from $n_e=n_0$ upstream to $n_{s}/n_{0} = (\Gamma_{\rm ad}/(\Gamma_{\rm ad}-1) + 1/\gamma_{0}(\Gamma_{\rm ad}-1)) \sim 3.1$\cite{Blandford_pof_1976,Kirk_npp_1999} downstream of the shock shown in Fig.~\ref{fig:density_fieldenergy}(d) corresponds to a hydrodynamic shock with an adiabatic constant $\Gamma_{\rm ad}=3/2$ of a 2D gas. On the $x>0$ side of the contact point $x=0$, the shock propagates with the velocity $v_{s}/c = (\Gamma_{\rm ad} - 1)((\gamma_{0} - 1)/(\gamma_{0} + 1))^{1/2} \sim 0.475c$\cite{Blandford_pof_1976,Kirk_npp_1999} in the $+x$-direction.

Effective collisions inside the shock are provided by the turbulent magnetic field plotted in Fig.~\ref{fig:density_fieldenergy}(b), where complex multi-filamentary structures with a typical transverse scale of $\sim 5 \, k_{p}^{-1}$ can be observed reaching from the shock into the upstream region. Magnetic filaments are elongated in the direction of the incoming upstream plasma. While the magnetic field is quasi-static in the down stream region, it is highly dynamic in the upstream region. Such time-dependence results in a finite longitudinal electric field $E_x$.

The relative magnitudes of different components of the electromagnetic field can be appreciated from the respective plots of their transversely averaged energy densities as shown in Fig.~\ref{fig:density_fieldenergy}(e). The largest normalized energy density $\epsilon_{B}(x) = \langle B^{2}_{y} \rangle /8 \pi \gamma_{0} n_{0} m_{e} c^{2}$ is associated with the magnetic field (black line), while the smallest one, $\epsilon_{x}^{(E)}(x) = \langle E^{2}_{x} \rangle /8 \pi \gamma_{0} n_{0} m_{e} c^{2}$, belongs to the longitudinal electric field (blue line). Here $\langle \rangle$ defines averaging over the transverse $z$ coordinate. The intermediate energy density $\epsilon_{z}^{(E)}(x) = \langle E^{2}_{z} \rangle /8  \pi \gamma_{0} n_{0} m_{e} c^{2}$ is associated with the transverse electric field (magenta line). Note that both $\epsilon_{B}(x)$ and $\epsilon_{z}^{(E)}(x)$ reach far into the upstream region, forming an important pre-shock region~\cite{Lemoine_prl_2019,Lemoine_pre_2019a,Lemoine_pr_2019b,Lemoine_pre_2019c} discussed below in the context of Fermi acceleration. In our simulation, the magnetic field energy density peaks at $\sim 20 \%$ of the equipartition energy in the shock transition region, and decays away from the shock front.

\subsection{The role of the longitudinal electric field in electron energy Maxwellization}\label{sec:long_field}

\begin{figure}[ht] \label{fig:momentum_bifurcation}
\includegraphics[width=1\linewidth]{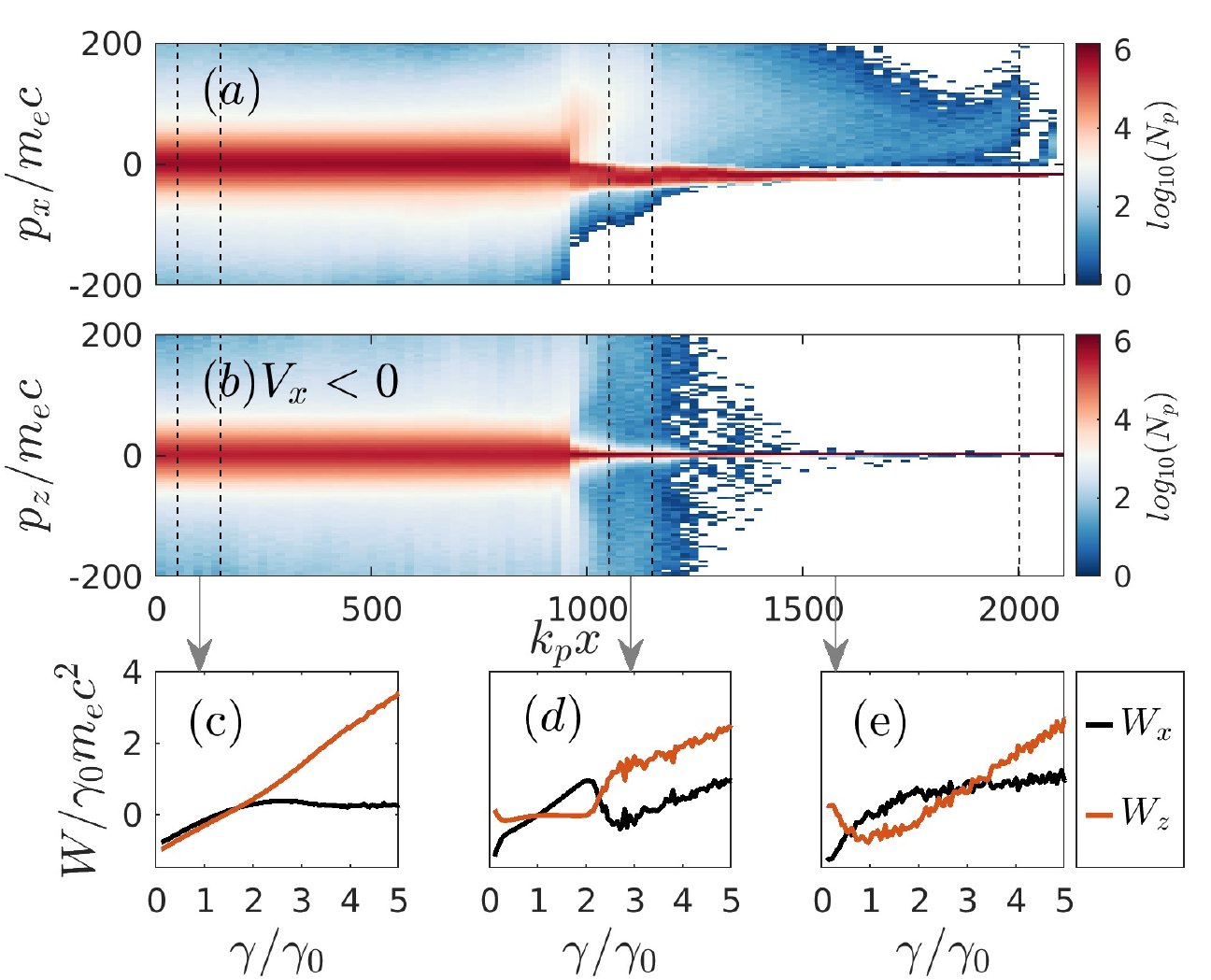}
\caption{Phase space densities in the downstream ($x<950 k_p^{-1}$), upstream ($x>1050 k_p^{-1}$), and shock/pre-shock ($950<x<1050 k_p^{-1}$) regions. (a) Longitudinal (all particles) and (b) transverse (counter-stream particles) color-coded momentum phase space density. (c-e) Decomposition of the kinetic energy gain/loss into the work done by longitudinal (blue line) and transverse (red line) electric fields inside selected spatial windows: (c) downstream ($50 <x< 150 k_p^{-1}$), (d) inside the shock ($1050<x<1150 k_p^{-1}$), and (e) upstream ($x>1150 k_p^{-1}$). }
\end{figure}

We further note from Fig.~\ref{fig:density_fieldenergy}(e) that the longitudinal electric field energy is vanishingly small in the upstream region, whereas the transverse electric and magnetic field energies are much stronger and comparable to each other: $\epsilon_{z}^{(E)} \approx \epsilon_B$ everywhere in the upstream region~\cite{Lemoine_prl_2019,Lemoine_pre_2019a}. The latter property is due to the fact that the dominant current filaments in the upstream (including the pre-shock) region are associated with highly-directional flows of electrons and positrons that have not yet undergone any significant isotropization as can be seen from Fig.~\ref{fig:momentum_bifurcation}(b). The presence of a small but finite longitudinal electric field in the shock transition region has been related to the oblique modes associated with the WI~\cite{Bret_PRE_2010,Lemoine_prl_2019}.

The role of the transverse component of the electric field in accelerating superthermal particles has been generally recognized~\cite{Lemoine_pre_2019c}. Its importance is not surprising because of its large amplitude in the pre-shock region. On the other hand, the role of the longitudinal electric field in providing energy Maxwellization to the medium-energy particles (both downstream and upstream) has not been previously studied. The reason for neglecting the $E_x$ component is that it is considerably smaller than $E_z$ in the pre-shock region. On the other hand, the amounts of the mechanical work $W_{x}^{(j)}$ ($W_{z}^{(j)}$) done by the longitudinal (transverse) electric field components on the $j$th upstream particle interacting with the shock field could be comparable with each other. Here we define
\begin{equation}\label{eq:work_done}
 W_{x,z}^{(j)} = q^{(j)} \int_{-\infty}^{+\infty}dt E_{x,z}(x^{(j)}(t),z^{(j)}(t)) v_{x,z}^{(j)}(t),
\end{equation}
where $v_{x,z}^{(j)}=p_{x,z}^{(j)}/m_{e}\gamma^{(j)}$ are the time-dependent longitudinal (transverse) velocity components of the $j$th particle. From here onwards, we assume that the particles are electrons, and $q^{(j)}=-e$.
We further note from Figs.~\ref{fig:momentum_bifurcation}(a,b) that most of the counter-streaming ($p_x<0$) particles are not yet isotropized, {\it i.e.}, $|p_z|^{(j)} \ll |p_x|^{(j)}$. This creates a surprising opportunity for $|W_x|^{(j)} \simeq |W_z|^{(j)}$ despite $|E_x|\ll |E_z|$ everywhere in the pre-shock region.

Moreover, we find that the two electric field energies, $\epsilon_{z}^{(E)}$ and $\epsilon_{x}^{(E)}$, are comparable in the downstream region, see the inset in Fig.~\ref{fig:density_fieldenergy}(e). The relativistic pair plasmas incident on the shock region are fully isotropized behind the shock, as can be observed by comparing Figs.~\ref{fig:momentum_bifurcation}(a) and (b). The out-of-plane magnetic field is responsible for efficient isotropization of the incident plasma: magnetic field energy $\epsilon_B$ dominates the downstream region immediately behind the shock, where it is much larger than the electric field energy. Therefore, just as in the pre-shock region, it is plausible for the two electric field components to do comparable mechanical work on the incident particles. In the next Section, we classify plasma electrons interacting with the shock into two categories defined by the relative magnitudes of $W_x$ and $W_z$.

\section{Emergence of the Energy Bifurcation}\label{bifurcation}
To quantify the contributions of the longitudinal and transverse electric fields to individual particles' kinetic energy increments $\Delta \varepsilon^{(j)} \equiv (\gamma^{(j)} - \gamma_0)m_{e}c^2$, where $\gamma^{(j)}$ is the final Lorentz factor of the $j$'th particle, we break up all particles located within a given spatial window $L_1 < x < L_2$ into energy bins centered around their final Lorentz factors $\gamma$. The following values of $L_1$ and $L_2$ indicated by the dashed lines in Fig.~\ref{fig:momentum_bifurcation}(a,b) were chosen: (i) $L_1=50 k^{-1}_{p}$ and $L_2 = 150 k^{-1}_{p}$ for the downstream region, (ii) $L_1= 1050k^{-1}_{p}$ and $L_2 = 1150 k^{-1}_{p}$ for the shock/pre-shock region, and (iii) $L_1= 1150 k^{-1}_{p}$ and $L_2 = 2000 k^{-1}_{p}$ for the upstream region. In order to concentrate specifically on the electrons that have already completed their interaction with, and thermalization by the shock, only those particles with $v_x^{(j)} > 0$ were counted in the upstream region. Note that a considerably larger spatial window was used for the upstream particles because of their relatively small number.

Inside each spatial window, an ensemble of particles whose final energy is centered around $\gamma$ is selected, and their average respective energy gains $W_{x,z}(\gamma) \equiv \langle W_{x,z}^{(j)} \rangle$ are calculated over the $\gamma$-dependent ensembles. The results for $W_{x,z}(\gamma)$ are plotted in Figs.~\ref{fig:momentum_bifurcation}(c-e) as a function of the final Lorentz factor $\gamma$ for the downstream (c), shock/pre-shock (d), and upstream (e) spatial windows. The most dramatic result corresponds to shocked plasma downstream from the shock: the $W_{x}(\gamma)$ (black) and $W_{z}(\gamma)$ (orange) curves plotted in Fig.~\ref{fig:momentum_bifurcation}(c) exhibit a clear bifurcation at $\gamma \equiv \gamma_{\rm bf} \approx 2 \gamma_0$.

Note that no such bifurcation was found for the electrons residing in the shock region, as can be seen from Fig.~\ref{fig:momentum_bifurcation}(d). We attribute this to the fact that the electrons inside the shock have not yet completed their interaction with turbulent electromagnetic fields inside and outside of the shock. Similarly, the small population of particles reflected by the shock back into the upstream region (see Fig.~\ref{fig:momentum_bifurcation}(e)) does not exhibit the same behavior of the $W_{x,z}(\gamma)$ graphs as the downstream particles. This behavior is discussed in Section~\ref{sec:reflected_particles}. Below we concentrate on the analysis of particle energy gain/loss in the downstream region.

\subsection{Properties of Thermalized Plasma Downstream From the Shock}\label{sec:downstream}

Based on the bifurcated curves in Fig.~\ref{fig:momentum_bifurcation}(c), we identify two groups of particles in the downstream region: particles with moderate ($\gamma < \gamma_{\rm bf}$) and particles with large ($\gamma > \gamma_{\rm bf}$) kinetic energies. The first group of particles, which we refer to as the bulk population, is thermalized to a relativistic Maxwellian distribution. Remarkably, both the longitudinal and transverse electric fields perform equal work on the bulk plasma particles: $W_{x}(\gamma) \approx W_{z}(\gamma)$ for all $\gamma < \gamma_{\rm bf}$. Note that the bulk population contains both particles that have been slowed down by the electric fields of the shock ($W_{x,z}(\gamma)<0$ for $\gamma < \gamma_0$) and the ones that have nearly doubled their energy. To our knowledge, this is the first computational demonstration of the equal contributions of the longitudinal and transverse components of the electric field in the Maxwellization of the shocked pair plasma. While the importance of the longitudinal field component $E_x$ has been known in electron-ion plasmas~\cite{Spitkovsky_ions_2008,Kumar_2015}, it has not yet been appreciated for collisionless shocks in pair plasmas~\cite{Lemoine_pr_2019b}.

The second group of particles, which we refer to as superthermal particles, acquire most of their kinetic energy from the transverse electric field, {\it i.e.}, $W_{z}(\gamma) > W_{x}(\gamma)$ for all $\gamma > \gamma_{\rm bf}$ as shown in Fig.~\ref{fig:momentum_bifurcation}(c). The bifurcation point at $\gamma = \gamma_{\rm bf}$ in the work-energy graph separates the population of the bulk particles gaining energy in the downstream region of the shock from the population of superthermal particles gaining energy in the course of repetitive bouncing in the shock/pre-shock region. By carrying out simulations for different periods of time, we have observed that while the ratio of the work performed by the transverse and longitudinal electric field increases with time for the superthermal population, the value of the bifurcation Lorentz factor $\gamma_{\rm bf}$ remains time-invariant.

\begin{figure}[ht] \label{fig:bifurcation}
\centering
\includegraphics[width=1\linewidth]{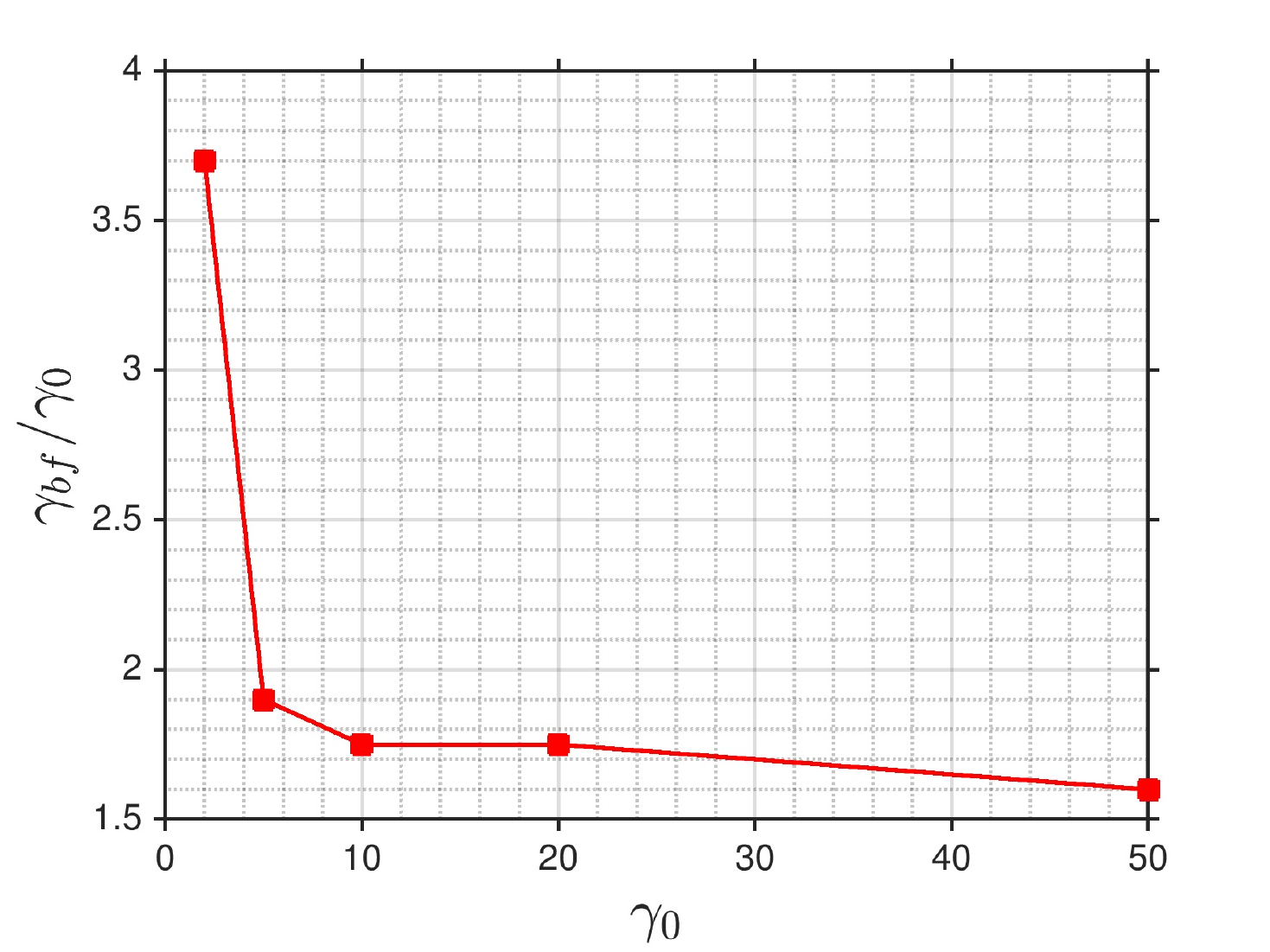}
\caption{The ratio of the bifurcation to initial (upstream) Lorentz factors $\gamma_{bf}/\gamma_0$ for different initial Lorentz factors $\gamma_0$.}
\end{figure}

We have also carried out simulations with varying initial (upstream) Lorentz factor ($\gamma_{0} = 2 - 50$). In Fig. \ref{fig:bifurcation} we have plotted the ratio $\gamma_{\rm bf}/\gamma_0$ as a function of $\gamma_0$. Remarkably, we found that for relativistic pair plasma shocks ($\gamma_{0} \geq 5$) the ratio $\gamma_{\rm bf}/\gamma_0$ remains constant ($ \approx 2$). However, for mildly relativistic shocks, this ratio is found to be much higher (e.g., $\approx 3.7$ for $\gamma_0 = 2$). It is expected that by symmetry, positrons and electrons exhibit the same bifurcating work-energy graphs.

\begin{figure}[ht] \label{fig:particle_spectrum}
\centering
\includegraphics[width=0.8\linewidth]{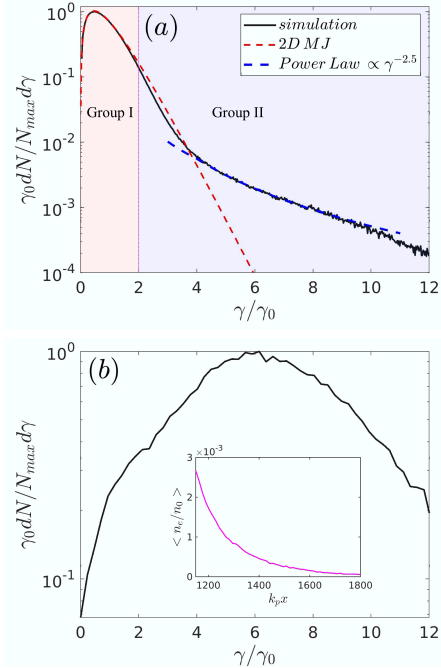}
\caption{(a) Electron energy spectra inside the $50 k_{p}^{-1} < x < 150 k_{p}^{-1}$ spatial window downstream: simulated (black line) and its fit to the sum of a Maxwell-J\"{u}ttner (red line) and a power-law $\gamma^{-2.5}$ (blue line) spectra. (b) Electron energy spectra of the upstream reflected particles inside the $1150 k_{p}^{-1} < x < 2000k_{p}^{-1}$ spatial window. Inset: transversely averaged density of upstream-reflected electrons at $t_{\rm fin} = 2040 \omega^{-1}_{p}$.}
\end{figure}

Another manifestation of the emergence of the superthermal population comes from the energy spectrum of thermalized electrons in the shocked region of the plasma. A typical spectrum plotted in Fig.~\ref{fig:particle_spectrum} corresponds to thermalized electrons inside a $100 \, k_{p}^{-1}$-wide slice in the downstream region at $t_{\rm fin} \simeq 2040 \omega^{-1}_{p}$. We have fitted the numerically simulated spectrum (black line) to a sum of a Maxwell-J\"{u}ttner (MJ)~\cite{juttner_AnnPhys1911} (red line) and a power-law (blue line) spectra. {Specifically, we chose the following analytic expression for the distribution function:
\begin{eqnarray} \nonumber \label{eq:distribution}
f \left( \frac{\gamma}{\gamma_0} \right) &  = & C_1 \frac{\gamma}{\gamma_0}  \exp \left( - \frac{\gamma}{ \Theta} \right) +  \\  && C_2 \left( \frac{\gamma}{\gamma_0} \right) ^{-p}
 min\left\lbrace 1, \exp \left( - \frac{\gamma - \gamma_{cut}}{\Delta \gamma_{cut}} \right) \right\rbrace,
\end{eqnarray}
where the first and second terms in the RHS correspond to the MJ and power law (with an exponential cutoff) \citep{Spitkovsky_2008,Stockem_ppcf_2012}  distributions, respectively.  The cutoff implies that $C_2 = 0$ for $\gamma < \gamma_{\rm min}$, $\Theta = k_{B}T/ m_{e}c^{2}$ is the dimensionless temperature, $k_{B}$ is the Boltzmann constant, and $C_{1,2}$ are the normalization constants. $\gamma_{\rm cut}$ and $\Delta \gamma_{\rm cut}$ show the beginning of high energy cutoff and high energy spread, respectively. In Fig. ~\ref{fig:particle_spectrum}(a), theoretical MJ distribution is plotted for $k_{B}T = 9.3 m_{e}c^{2}$, which is in excellent agreement with $k_ {B}T_{\rm RH} = 0.5 (\gamma_0 -1)m_{e}c^{2} = 9.5m_{e}c^{2}$ predicted by Rankine-Hugoniot condition \cite{Blandford_pof_1976,Kirk_npp_1999}  for complete thermalization in the downstream region. The power-law is plotted for $p = 2.5$, $\gamma_{\rm min} = 3 \gamma_{0}$, $\gamma_{\rm cut} = 10 \gamma_{0}$ and $\Delta \gamma_{\rm cut} = 6 \gamma_{0}$.} Deviation from the MJ spectrum is clearly observed for $\gamma > 2\gamma_0$. Understanding the origins of the two groups of particles (bulk (group I) and superthermal (group II)) requires that we examine individual particle trajectories in greater detail: their entrance into the shock, subsequent interaction with the shock, and their eventual transition into the downstream region.

\subsection{Particle Tracking Results}\label{single_particle_dynamics}
\begin{figure}[ht]\label{fig:four_particle_trajectory}
\includegraphics[width=1\linewidth]{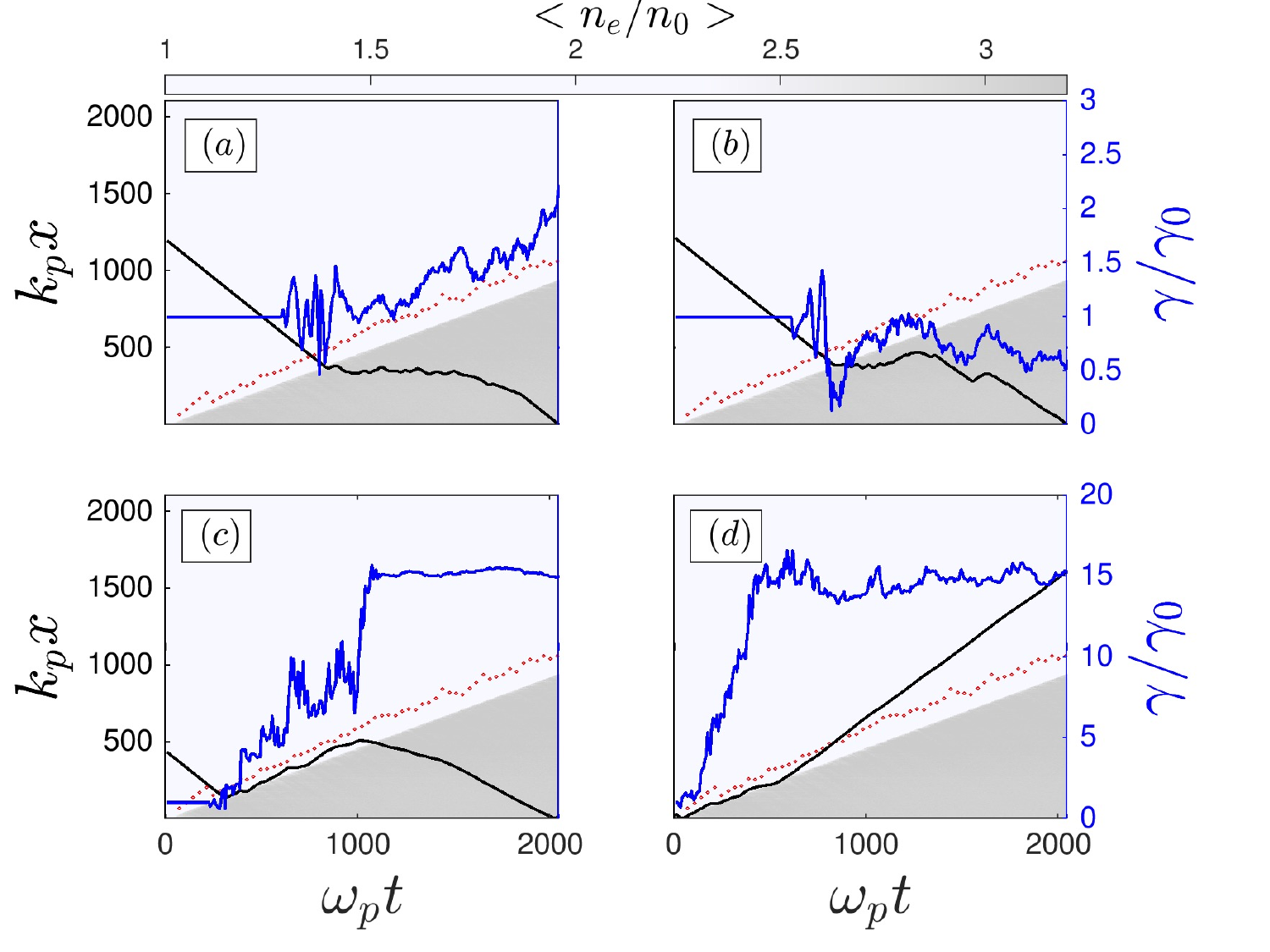}
\caption{Time-dependent trajectories and normalized energies of four representative electrons. Color-coded: transversely-averaged electron density separated by the shock line $x_{\rm sh}(t)$. Dotted red line: the boundary of the pre-shock $x_{\rm p-sh}(t)$. Black lines (left scale): horizontal trajectories $x(t)$, blue lines (right scale): energies $\gamma(t)$. (a,b) Typical bulk electrons gaining (a) and losing (b) energy. (c,d) Superthermal electrons moving into downstream (c) and upstream (d) regions.}
\end{figure}

To understand field-particle interactions with different regions of the shock, we tracked electrons' normalized energies $\gamma(t)$ and positions $x(t)$ between $t \equiv t_0=0$ and $t \equiv t_{\rm fin}=2040 \omega^{-1}_{p}$ based on their final energies $\gamma_{\rm fin} \equiv \gamma(t_{\rm fin})$ and positions $x_{\rm fin} \equiv x(t_{\rm fin})$ with respect to the shock. Specifically, four classes of particles were considered: (I) two classes of the bulk particles with $\gamma_{\rm fin} < \gamma_{\rm bf}$ that ended up downstream of the shock's position $x_{\rm sh}(t) \approx v_{\rm sh}t$ (top row of Fig. \ref{fig:four_particle_trajectory}), and (II) two classes of superthermal particles with $\gamma_{\rm fin} \gg \gamma_{\rm bf}$ (bottom row of Fig. \ref{fig:four_particle_trajectory}). Group I electrons comprise those that gained or lost energy, as exemplified by representative particles in Figs. \ref{fig:four_particle_trajectory}(a) and (b), respectively. Group II electrons that gained a significant amount of energy from the shock comprise those that have crossed the shock into the downstream region, as shown in Fig.~\ref{fig:four_particle_trajectory}(c), and those that have reflected from the shock into the upstream region, as shown in Fig.~\ref{fig:four_particle_trajectory}(d).

The black lines in Fig.~\ref{fig:four_particle_trajectory} indicate electrons' trajectories $x(t)$. Shock's trajectory $x_{\rm sh}(t)$ separates the blue (upstream) region from the gray (downstream) region. The dotted red line indicates the edge of the pre-shock region $x_{\rm p-sh}(t)$ defined in such a way that the magnetic energy declines from its peak as $\epsilon_B(x_{\rm p-sh})/\epsilon_B(x_{\rm sh})=1/e^2$ (see Fig.~\ref{fig:density_fieldenergy}(e) for a representative profile of $\epsilon_B$ as a function of $x$). The blue lines in Figs.~\ref{fig:four_particle_trajectory}(a-d) indicate electrons' Lorentz factors $\gamma(t)$ normalized by $\gamma_0$, i.e. energy gain (loss) correspond to $\gamma_{\rm fin}/\gamma_0 > 1$ ($\gamma_{\rm fin}/\gamma_0 < 1$). By comparing particles' trajectories and energy changes, it is easy to deduce when those energy changes have occurred.

Particles of the first and second classes of group I gain or lose moderate amounts of energy that are comparable to their initial energies $\varepsilon_0 = \gamma_0 m_{e}c^2$. Those particles cross the shock once, become thermalized, and turn into bulk plasma in the downstream region. The downstream plasma primarily consists of these two classes of particles, as they form the Maxwellian portion of the spectrum shown in Fig.~\ref{fig:particle_spectrum}. As the bifurcation in Fig.~\ref{fig:momentum_bifurcation}(c) indicates, these two classes of particles, on average, gain (for $W_{x,z}>0$) or lose (for $W_{x,z}<0$) approximately equal amounts of energy from both components of the electric field. This is related to the fact that the downstream region of the plasma contains almost equal amounts of electromagnetic energies $\epsilon_{x}^{(E)}$ and $\epsilon_{z}^{(E)}$ associated with the longitudinal and transverse electric field components, respectively (see inset in Fig.~\ref{fig:density_fieldenergy}(e)). Additional mixing between longitudinal and transverse momenta $p_x$ and $p_z$ is provided by the magnetic field $B_y$ which is much larger than either $E_x$ or $E_z$ components of the electric field.

A small number of particles which are either reflected by the shock, or diffuse from the downstream to upstream region, do not immediately cross the high-field region between the shock and the pre-shock boundary shown by a dashed line in Figs.~\ref{fig:four_particle_trajectory}(c) and (d). Such group II particles can stay in the pre-shock region for a long time, gaining significant energy from the strong transverse electric field. An example of a particle belonging to the third class of bulk electrons that gain considerable energy while eventually moving through the shock is shown in Fig.~\ref{fig:four_particle_trajectory}(c). This specific particle (which we label as $j=3$) stays in the pre-shock region for almost $(\Delta t)^{(3)}_{\rm p-sh} \approx 700\omega_p^{-1}$, gains $\Delta \varepsilon^{(3)} \approx 14 \varepsilon_0$ by experiencing numerous rapid energy changes that can be characterized as first-order Fermi accelerations, and eventually crosses the shock transition region into the downstream region.

In agreement with the energy bifurcation curve, $W_z^{(3)} \approx 12 W_x^{(3)}$, {\it i.e.}, superthermal electrons crossing into the downstream region gain more than an order of magnitude from the transverse component of the electric field than from the longitudinal one. The reason for this is that superthermal electrons spend a long period of time $(\Delta t)^{(3)}_{\rm p-sh}$ in the pre-shock region, where they are subjected to $E_z \gg E_x$. In combination with isotropization provided by a strong magnetic field in the pre-shock region, this results in $W_z^{(3)} \gg W_x^{(3)}$.

\section{Reflected particles in the upstream region}\label{sec:reflected_particles}
At the same time, a minority of electrons interacting with the pre-shock region for a long time eventually get reflected and move into the upstream region. The number of reflected electrons and positrons is much smaller than of those propagating past the shock into the downstream region. Qualitatively, this is related to the fact that the combination of the transverse magnetic and electric fields in the pre-shock region creates a stronger deflecting force for the particles traveling in the positive $x$-direction than for their counterparts with $v_x<0$. Therefore, the pre-shock creates an effective one-way barrier that makes it easier for the thermalized particles to diffuse downstream from the shock than to reflect back into the upstream region.

The energy spectrum of the reflected electrons population is shown in Fig.~\ref{fig:particle_spectrum}(b). It peaks at a much higher Lorentz factor $\gamma^{({\rm up})}_{({\rm peak})} \approx 5\gamma_0$ than the $\gamma^{({\rm down})}_{({\rm peak})} \approx \gamma_0$ peak of the energy spectrum of the downstream electron population. Therefore, based on the plots of the averaged $W_x(\gamma)$ and $W_z(\gamma)$ in Fig.~\ref{fig:momentum_bifurcation}(e), we conclude that most of the reflected pairs gain most of their energy from the transverse electric field component than from the longitudinal one. A typical trajectory and energy gain plots for a representative class-four particle are shown in Fig. \ref{fig:four_particle_trajectory}(d). The particle spends roughly the same time interacting with the pre-shock as the one shown in Fig.~\ref{fig:four_particle_trajectory}(c), gains approximately the same energy, and eventually becomes a counter-streaming particle penetrating deep into the upstream region.

Next, we discuss the importance of the counter-streaming particles for seeding the WI. The counter-streaming population propagating ahead of the shock, plotted in the inset of Fig. ~\ref{fig:particle_spectrum}(b) and also observed in Fig.~\ref{fig:momentum_bifurcation}(a), is essential for maintaining the shock. For example, the density of counter-streaming particles determines the growth rate and saturation of the secondary WI manifested magnetic field filaments in the upstream region, as shown in the Fig.~\ref{fig:density_fieldenergy}(d).

\begin{figure}[ht]\label{fig:gamma_density}
\includegraphics[width=1\linewidth]{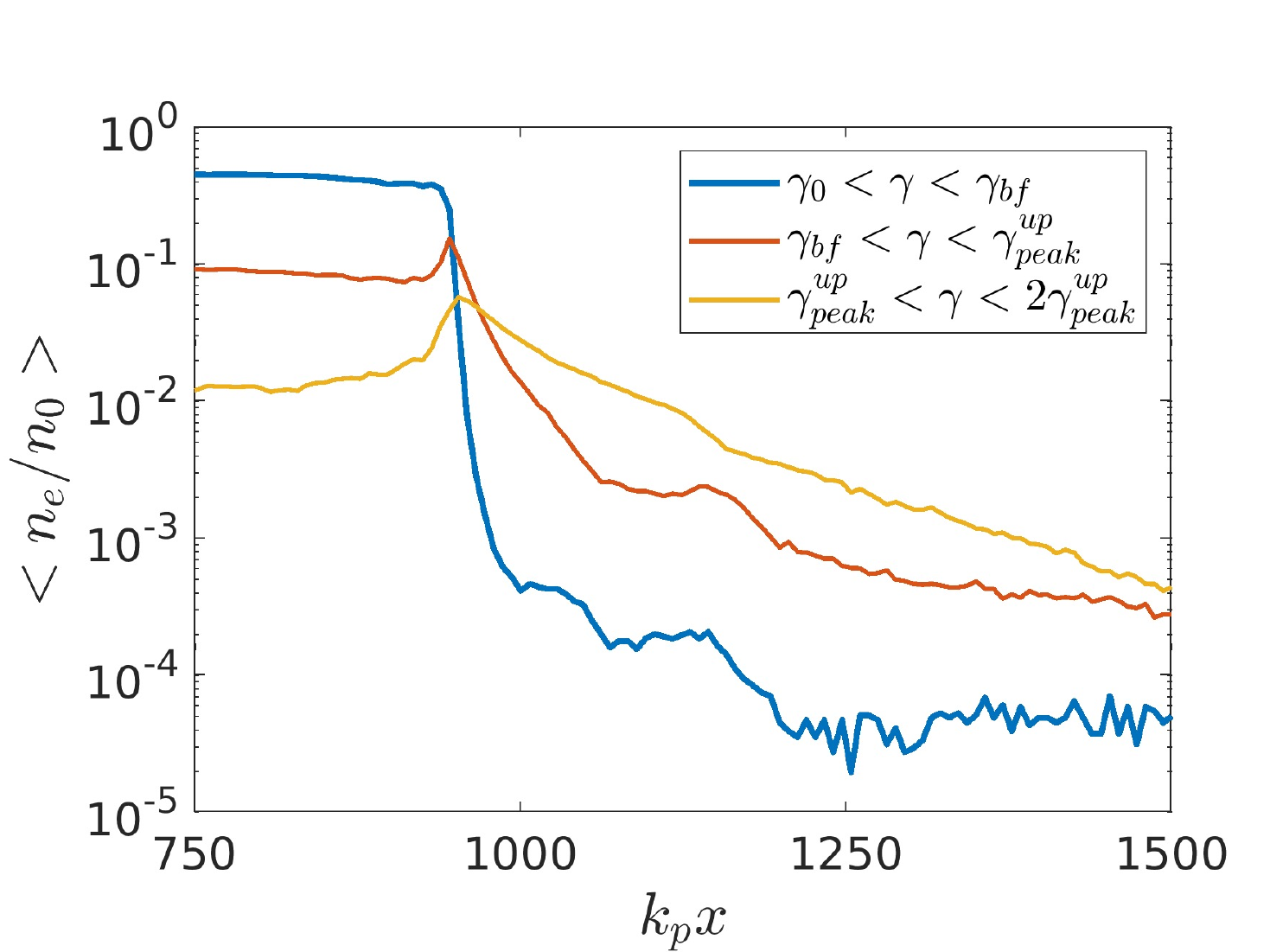}
\caption{Transversely averaged electron densities for kinetic energy ranges  $\gamma_0 < \gamma < \gamma_{\rm bf}$ (blue line), $\gamma_{\rm bf} < \gamma < \gamma^{({\rm up})}_{({\rm peak})}$ (orange line), and $\gamma^{({\rm up})}_{({\rm peak})} < \gamma < 2 \gamma^{({\rm up})}_{({\rm peak})}$ (yellow line). All plots correspond to $t=t_{\rm fin}$.}
\end{figure}

Note that the density of the counter-streaming particles decreases as they move away from the shock transition region. This effect is illustrated by Fig.~\ref{fig:gamma_density}, where we plot the transversely-averaged density of electrons with Lorentz factors within the following ranges: (1) $\gamma_0 < \gamma < \gamma_{\rm bf}$ (blue line), (2) $\gamma_{\rm bf} < \gamma < \gamma^{({\rm up})}_{({\rm peak})}$ (orange line), and (3) $\gamma^{({\rm up})}_{({\rm peak})} < \gamma < 2 \gamma^{({\rm up})}_{({\rm peak})}$ (yellow line).

The general trend is the same for all three energy ranges: higher density in the downstream than in the upstream region. It confirms that the particles more easily escape into the downstream than into the upstream because of the deep penetration of the transverse electric and magnetic fields into the upstream region shown in Fig.~\ref{fig:density_fieldenergy}(e). Only the highest energy highly-collimated counter-streaming particles penetrate deep into the upstream as they are less scattered by the upstream electromagnetic fields, which clearly explains rapid density fall of the counter-streaming particles in the upstream region shown in the inset of Fig. \ref{fig:particle_spectrum}(b). Another reason behind the lower density of accelerated particles in the upstream is that the reflected particles seed the secondary Weibel instability in the upstream region, thereby losing energy in the process~\cite{Lemoine_pre_2019a}. As more particles are accelerated by the shock, the resulting sub-population of fast particles catches up with the slower particle reflected at earlier times. This leads to overall density increase of counter-streaming particles with time, which is likely to be the reason why current PIC simulations do not reach a steady-state~\cite{Keshet_apj_2009}.


\begin{figure}[ht]{ \label{fig:trajectory_single_particle}}
\centering
\includegraphics[width=1\linewidth]{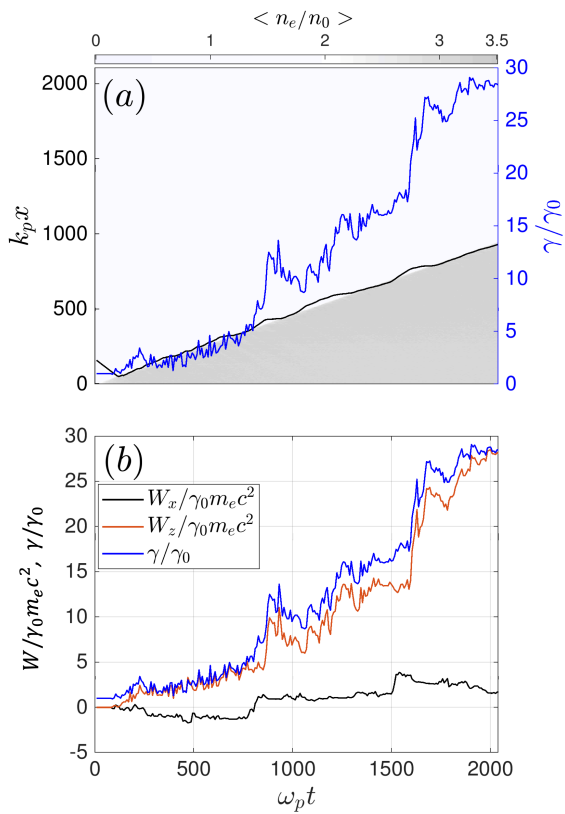}
\caption{Time evolution of the longitudinal position and normalized energy of a representative fast electron reflected by the shock. (a) Black line: particle trajectory $x(t)$, blue line:  Lorentz factor $\gamma(t)$. Color-coded: transversely-averaged plasma density. (b) Mechanical work performed by the longitudinal (black line) and transverse (orange line) electric field components, and the Lorentz factor (blue line) of the particle.}
\end{figure}

Not surprisingly, some of the most energetic electrons can be found among those reflected upstream of the shock. The trajectory of one such simulated particle shown in Fig.\ref{fig:trajectory_single_particle}(a) (blue line, left scale) demonstrates that the most energetic class-four particles ``surf''  around the shock and gain energy continuously (blue line, right scale). The temporal evolution of the decompositions of the kinetic energy change into $W_{x}^{(4)}(t)$ and $W_{z}^{(4)}(t)$ are plotted in Fig.~\ref{fig:trajectory_single_particle}(b).  The jumps in $W_{z}^{(4)}(t)$ clearly coincide with multiple scatterings of the particle around the shock region. Such scattering in the shock transition region can be identified as a beginning of the first order Fermi acceleration.

\section{Conclusions} \label{conclusions}
In conclusion, we have studied particle acceleration via the unmagnetized relativistic collisionless shock in pair (electron-positron) plasmas by means of a first-principles 2D PIC code. The vectorial nature and strong anisotropy of the electric field produced by the classic Weibel instability contributes to highly anisotropic energy gain by fast particles experiencing first order Fermi acceleration in the shock and pre-shock regions. On the other hand, the electric field is found to be fairly isotropic in the shocked plasma region downstream of the shock. The effects of the anisotropic electric fields on the particles upstream and downstream of the shock were studied by implementing a particle tracking routine that follows, as a function of time, the mechanical work done by each field component on the individual particles.

One of the key findings of tracking particles' energy gains and losses is that the downstream plasma particles bifurcate into two groups based on their final energy: slow ($\gamma < \gamma_{\rm bf}$) and fast ($\gamma > \gamma_{\rm bf}$) group of particles. For relativistic shocks, the empirically found value of the bifurcation Lorentz factor separating the two groups is found to be $\gamma_{\rm bf} \sim 2 \gamma_{0}$ for a wide range of the initial Lorentz factors $\gamma_0$. Another surprising findings of particle-tracking is that the slow group of particles forming the bulk of the shocked plasma gains/loses equal amounts of energy from the longitudinal and transverse electric field components despite the former being much smaller than the latter in and around the shock. On the other hand, the fast particles gain most of their energy from the transverse electric field component because most of the energy gain takes place inside the shock/pre-shock region, where particles' momenta are already thoroughly anisotropized while the longitudinal component of the electric field is much smaller than the transverse one. Therefore, the results of tracking particles' trajectories and energy exchanges with the two electric field components indicate that the development of a bifuracated energy gain distribution is a telltale sign of the emergence of Fermi acceleration in the shock/pre-shock regions of the plasma. Future research directions will include extending these results to 3D geometry, as well as expanding this work to mixed plasma flows containing hadrons in addition to leptons.

\begin{acknowledgments}
The work was supported by DOE grant DE-NA0003879. The authors thank the Texas Advanced Computing Center (TACC) for providing HPC resources.
\end{acknowledgments}
\vspace{5mm}
\section*{data  availibility statement}
The data that support the findings of this study are available from the corresponding author upon reasonable request.

\bibliography{work_bifurcation_GS_v3.bib}
\end{document}